\documentclass{aa}  
\usepackage{graphicx}
\usepackage{txfonts}
\begin{document}
%
   \title{The reddest ISO--2MASS quasar
           \thanks{Based on observations made
           with ESO Telescopes at La Silla and Paranal under programme IDs
           275.A-5064 and 075.A-0374, with the Spitzer Space
           Telescope, which is operated by the 
	   JPL, CALTECH under a contract with NASA, and obtained at
           CTIO a division of NOAO, which is operated by AURA under
           cooperative agreement with the National Science
           Foundation.} 
   }




   \author{C. Leipski
          \inst{1,}\thanks{\emph{Present address:} University of California, Santa Barbara, \mbox{CA--93106}, USA; leipski@physics.ucsb.edu}
          \and
          M. Haas
          \inst{1}
          \and
          R. Siebenmorgen
          \inst{2}
          \and
          H. Meusinger
          \inst{3}
           \and
          M. Albrecht
          \inst{4}
          \and
          C. Cesarsky
          \inst{2}
          \and\\
          R. Chini
          \inst{1}
          \and
          R. Cutri
          \inst{5}
	  \and
	  H. Drass
	  \inst{1}
          \and
          J. P. Huchra
          \inst{6}
	  \and
          S. Ott
          \inst{7}
          \and
          B. J. Wilkes
          \inst{6}
          }


   \institute{Astronomisches Institut Ruhr--Universit\"at Bochum (AIRUB),
              Universit\"atsstra{\ss}e 150, 44780 Bochum, Germany
             \and
             European Southern Observatory (ESO),
             Karl--Schwarzschild--Str. 2, 85748 Garching, Germany
            \and
              Th\"uringer Landessternwarte Tautenburg (TLS), Sternwarte 5,
             07778 Tautenburg, Germany
             \and
             Instituto de Astronom\'ia, Universidad Cat\'olica del
             Norte (UCN),  Avenida Angamos 0610, Antofagasta, Chile
             \and
             IPAC, California Institute of Technology
             (Caltech),  770 South Wilson Avenue, Pasadena, CA
              91125, USA
             \and
             Harvard--Smithsonian Center for Astrophysics (CfA), 60
              Garden Street, Cambridge, MA 02138, USA
              \and
             HERSCHEL Science Centre, ESA, Noordwijk, PO Box 299, 2200
             AG Noordwijk, The Netherlands
   }

   \date{Received August 30, 2006; accepted July 11, 2007}

  \abstract 
{While there is growing consensus on the existence of
  numerous dust--enshrouded red quasars, their discovery and detailed
  exploration is still an observational challenge.  In the course of
  the near--mid--infrared AGN search combining the 6.7 $\mu$m ISOCAM
  Parallel Survey and 2MASS we have discovered 24 type\,--1 quasars
  about a third of which are too red to be discriminated by optical/UV
  search techniques.}  
{Here we report on
  a detailed case study of the reddest type\,--1 quasar of our sample,
  2MASS J23410667$-$0914327 (for short J2341)
  at redshift $z=0.236$ with M$_{K}=-25.8$ and $J-K_s=1.95$.
  This source has a 
    very red optical appearance but lacks the 
    far-infrared emission typically seen in known dust enshrouded AGN. 
    Therefore we here explore its enigmatic nature.}
{We performed spectroscopy in the optical with VLT/FORS1 and in the
  mid--infrared ($5-38\,\mu$m) with the {{\it Spitzer Space Telescope}} as
  well as near--infrared (NIR) imaging with ISPI at the CTIO
  4m--telescope. To explain the red optical continuum of the quasar we
  examine nuclear dust reddening of an intrinsically blue quasar continuum
  in combination with dilution by stellar light of the host galaxy. }  
{The optical and NIR observations reveal a  star forming emission--line galaxy 
at the same redshift as the quasar with a projected linear
  separation of 1\farcs8 (6.7 kpc).  The quasar and its companion are
  embedded in diffuse extended continuum emission. 
Compared with its companion the quasar exhibits redder optical--NIR 
colours,  which we attribute to hot nuclear dust.
  The mid--infrared spectrum shows only few
  emission lines superimposed on a power--law spectral energy
  distribution typically seen in type\,--1 AGN. However, the lack of
  strong far--infrared emission suggests that our potentially interacting
  object contains much less gas and dust and is in a stage
  different from dust reddened ULIRG--AGN like Mrk\,231.
  The optical spectrum shows signatures for reddening in the
  emission--lines and no post--starburst stellar population is detected 
  in the host galaxy of the quasar. The optical continuum emission of the active nucleus appears
  absorbed and diluted. } 
{   Even the combination of absorption and host dilution is not
    able to match J2341 with standard quasar templates.
    While the BLR shows only a rather moderate
    absorption of E$_{B-V}=0.3$, the continuum shorter than $4500\,\AA$
    requires strong obscuration with E$_{B-V}=0.7$, clearly exceeding the 
    constraints
    from the low upper limit on the silicate 9.7\,$\mu$m absorption.
    This leads us to conclude that the continuum of the quasar J2341
    is intrinsically redder than that of typical quasars.
}


   \maketitle
%

\section{Introduction}

  Pure optical quasar surveys find QSOs with essentially blue continua
  (e.g. Schmidt \& Green \cite{schmidt83}; Wolf \cite{wolf05}).
  However, investigations including the radio, X--ray  and infrared
  have revealed a substantial fraction of dust--reddened AGN missed by
  optical surveys (Low et
  al. \cite{low88}; Gregg et al. \cite{gregg02}; Maiolino et
  al. \cite{maiolino03}; Glikman et al. \cite{glikman04}; Lacy et al. \cite{lacy04}).
  We have combined  the ISOCAM Parallel Survey at
  6.7 $\mu$m (Siebenmorgen et al. \cite{siebenmorgen96}) with the
  2MASS survey (Skrutskie et al. \cite{skrutskie06}) in order to  obtain a
  powerful tool to search for AGN by means of their infrared colours 
 ($H-K_s > 0.5$ and $K_s-LW2(6.7\,{\mu{\rm m}}) > 2.7$; Haas et al. \cite{haas04}). In
  fact, in a high galactic latitude area of $\sim$\,10 deg$^2$ we found
  $30$\% more type\,--1 quasars per square degree down to $R$\,$=$\,18\,mag
  than e.g. the SDSS DR3 quasar survey (Leipski et
  al. \cite{leipski05}).  The quasars found by our NIR/MIR method show
  a variety of spectral shapes, in particular in the optical. Some of
  the ISO--2MASS objects are too red to be recognised as quasars by
  current optical strategies, probably because of dust extinction.
  Since the NIR and MIR colours of quasars are different from those of
  stars, infrared selection enables the detection of optically red
  quasars (Cutri et al. \cite{cutri02}; Smith et al. \cite{smith03}; Leipski et
  al. \cite{leipski05}; Lacy et al. \cite{lacy07}). 
  
  From our sample of 77 AGN candidates 24 turned out to be type\,--1
  AGN.
They span a range of optical colours from blue quasar--typical ones 
to very red ones similar to those found in dust--enshrouded
ULIRG-AGN.   
The reddest quasar of our sample is 2MASS J23410667$-$0914327 
(hereafter called J2341) at a moderate redshift $z=0.236$. 
Thus, its red photometric appearance is not due to high redshift. 

  Alternatively to the classical M$_{B}$\,$<$\,$-23$\,mag quasar definition,
  its absolute $K_s$--band brightness of M$_{K_s}$\,=\,$-25.8$\,mag
  qualifies J2341 as a QSO. 
  The absolute SDSS i magnitude of M$_i=-23.0$\,mag
  ~and an 
  [\ion{O}{iii}]\,$\lambda$5007 luminosity of
  ${\rm L}_{[\ion{O}{iii}]}$\,$\sim$\,$2.7\times10^{41}$~erg\,s$^{-1}$
  places J2341 at the low luminosity end of the total luminosity distribution
  of quasars found in the SDSS, but at intermediate values for objects in the
  $z=0.2-0.3$ regime (Schneider et al. \cite{schneider07}). 
  While no radio counterpart is listed by
  NVSS (translating to $F_{1.4\,{\rm GHz}}<2.5$\,mJy; Condon et al. 
  al. \cite{condon98}), ROSAT HRI   
  observations give a soft X--ray luminosity of  
  ${\rm L}_{0.5-2.4\,{\rm keV}}\sim3\times10^{43}$~erg\,s$^{-1}$,
  similar to that of low redshift PG--quasars. The galactic foreground 
  extinction is E$_{B-V}=0.03$ (Schlegel et al. \cite{schlegel98}).

 The low IRAS 60\,$\mu$m upper limit $F_{60\,\mu{\rm m}}<129$\,mJy
 suggests that J2341 is different from known ULIRG-AGN. 
 If this is true, then its red optical colours are puzzling. 
 Therefore we performed a detailed case study of J2341, using 
 optical VLT spectra, near--infrared 
 imaging from the CTIO 4m telescope and mid--infrared spectra from 
 the {\it Spitzer Space Telescope}.

\section{Observations}

\begin{figure}
  \resizebox{\hsize}{!}{\includegraphics[angle=0]{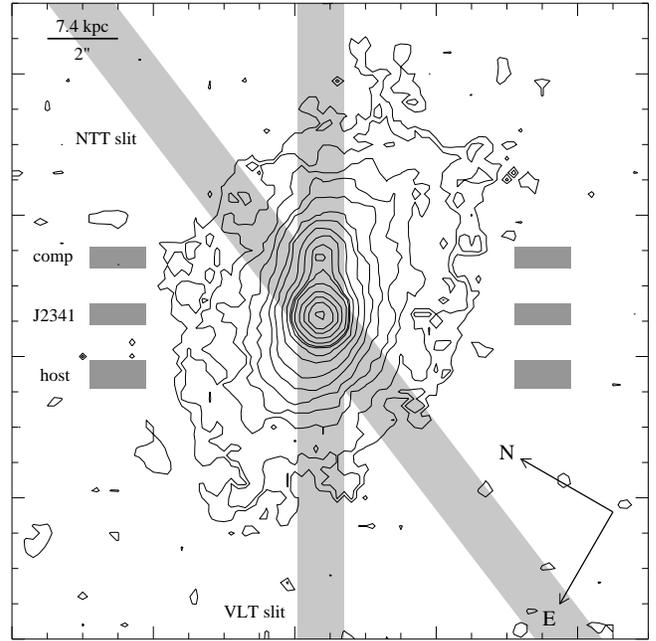}}
  \caption{     Contour plot of the J2341 ``system'' from the 30\,sec VLT $R$--band
  acquisition image. The contour levels have been chosen to highlight the extended
  emission. The image shows 18\arcsec\,$\times$\,18\arcsec~(90\,px\,$\times$\,90\,px). The positions
    and widths of the VLT slit (p.a. 120\degr; 1\farcs3 width) and the NTT slit
    (p.a. 155\degr; 1\farcs5 width) are indicated by the lightly
    shaded rectangles. The darker shaded areas indicate the positions and
    widths of the extraction regions for the spectra used in this paper.
  \label{acq_with_contours}
  }
\end{figure}

 The quasar nature of J2341 was discovered on spectra taken with
 EMMI at the ESO/NTT telescope during the optical follow--up spectroscopy 
of our MIR selected AGN candidates. We used a 1\farcs5 wide long slit  
oriented along the parallactic angle (slit--loss effects due to atmospheric
dispersion are minimised by such a slit orientation) and centred on the 
peak of the apparent brightness distribution (Fig.\,\ref{acq_with_contours}). These 
observations were obtained before the object was recognised as a double
source.  The slit
width in combination with Grism\#2 results in a spectral resolution of 
$\sim$\,$600\,$km\,s$^{-1}$. The integration time was $3\times20\,$min.

In order to check morphological details and the NIR photometry, 
we have obtained $JHK_s$ images with ISPI at the CTIO 4m--Blanco
 telescope. These observations were performed under stable seeing
 conditions (FWHM $\sim$ 1$\arcsec$) with a total on--source integration time of 15
 min in each filter. The images were reduced using standard IRAF procedures. 

After detecting that J2341 may be a double source, we 
 have further obtained optical spectra with VLT/FORS1 in order to explore the 
 nature of J2341 and its putative companion.  
The 1\farcs3 wide slit was 
 oriented to cover both objects (Fig.\,\ref{acq_with_contours})
 and an atmospheric dispersion corrector was
 used for the $1\times15\,$min observation. 
In combination with grism GRIS\_300V the observations yield 
a spectral resolution of $\sim$\,$650\,$km\,s$^{-1}$.

The optical spectra were reduced using standard procedures within the ESO/MIDAS
package. Since no spatial structure could be identified on the 2D
spectroscopic frames from the NTT run we combined all the flux into a single
 spectrum. This procedure may include contributions from the
 underlying host galaxy and from the companion, diluting the QSO signatures in the
 final NTT spectrum. 
The spectroscopic observations with the VLT were able to spatially resolve
both objects since the slit was optimally oriented. We searched for the 
brightest spectral column in broad line flux and in continuum for the QSO and
 the companion, respectively. We extracted 3--pixel wide columns
 (0.2\arcsec/px) into single
 spectra, each corresponding to only
 0\farcs6 spatially. Since the seeing was $\sim$\,1\farcs0 during the
 observations this choice reduces the flux in the extracted spectra but minimises the
 contributions from the adjacent source while providing adequate S/N ratios. 

 We also obtained a low--resolution $5-38\,\mu$m MIR spectrum of
 J2341 using IRS (Houck et
 al. \cite{houck04}) aboard {\it Spitzer Space Telescope} (Werner et
 al. \cite{werner04}) with total integration times of 240\,sec and
 480\,sec for SL and LL, respectively. We used data processed by 
 version S15.3.0 of the IRS pipeline and performed calibration and
 extraction of the spectra within SPICE.


\section{Comparison of the NTT and VLT spectra}

Here we briefly consider whether observational effects or 
host galaxy contamination is important for
the red appearance of the quasar spectrum and how far the quasar and
companion spectra may be contaminated by each other. 

\begin{figure}
  \resizebox{\hsize}{!}{\includegraphics[angle=0]{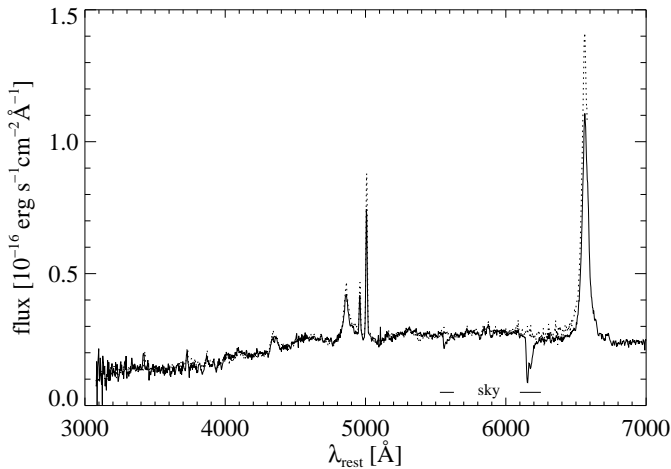}}
  \caption{Comparison of the VLT spectrum (dotted line) and the 
 NTT spectrum (solid line; scaled to match the continuum between 4000 and 6000 $\AA$).  
  \label{vlt_ntt}
  }
\end{figure}

In Fig.\,\ref{vlt_ntt} we plot the NTT spectrum (extracted through a
large aperture) over the VLT spectrum of the QSO (extracted through a
narrow aperture). The shape of the continuum
agrees very well for both observations. We conclude that
the red continuum is intrinsic to the source and not an artifact from the
data reduction or due to slit--loss effects. In addition, the red
optical continuum is also evident from USNO and SDSS photometry 
(e.g. Fig.\,3 in Leipski et al.\,\cite{leipski05}).

Compared with the VLT spectrum, 
the NTT spectrum includes more flux from the underlying host
galaxy and potentially some flux from the companion as well. 
Therefore it
shows more pronounced and even additional stellar absorption
lines. The increased continuum flux also results in smaller equivalent
widths (EWs) of the emission lines (Fig.\,\ref{vlt_ntt}).
  
Despite the stronger stellar contribution in the NTT spectrum,
the shape of the continuum is the same for both the VLT and NTT
spectra. This suggests that the influence of stellar light
contributions on the shape of the observed continuum can be significant
even for the nuclear VLT spectrum. We  will address the underlying
host galaxy further in Section\,\ref{section_dilution_by_host}.

In Figs.\,\ref{acq_with_spectra} and \ref{optical_zoomed}  
we show the
VLT spectra of the QSO and its companion. These spectra were extracted
through narrow apertures to minimise contaminations from each other. 
The absence of broad emission--line components
and emission from the highly ionised inner parts of the
narrow--line region (e.g. [\ion{Ne}{v}]) in the spectrum of the
companion strongly argues against significant contributions 
from emission of the QSO. The QSO
spectrum does not show signs for higher order Balmer absorption
lines which are on the other hand prominent in the companion. 
This supports our conclusion that the
spectra of the QSO and the companion show the intrinsic features of the
respective source and that they are not significantly contaminated by 
each other.


\section{Results and Discussion}

\begin{figure}
  \resizebox{\hsize}{!}{\includegraphics[angle=0]{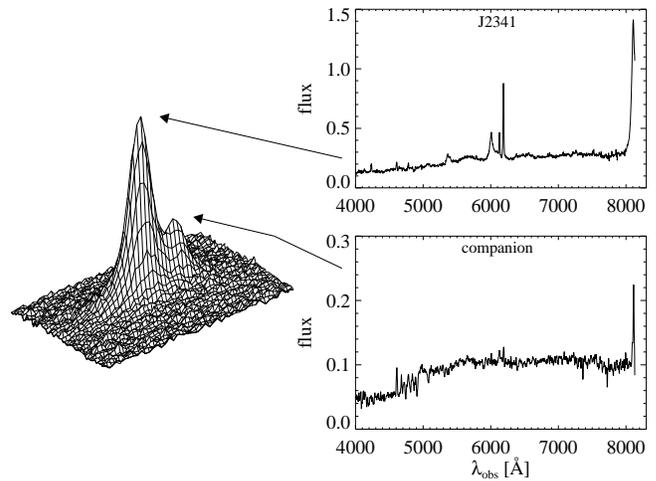}}
  \caption{$R$--band surface plot and spectra of the two objects
    obtained with FORS1/VLT. The flux is 
    given in 10$^{-16}$\,erg\,sec$^{-1}$\,cm$^{-2}$\,\AA$^{-1}$.
    \label{acq_with_spectra}
  }
\end{figure}

The inspection of the $JHK_s$ frames and the $R$--band acquisition
image (30\,sec integration time) reveals a nearby faint object approximately 1$\farcs$8
north--west of J2341 (Figs.\,\ref{acq_with_contours} and \ref{acq_with_spectra};
position angle $\sim$\,294 deg), which could be a physically interacting companion
galaxy. 
The VLT spectra reveal that both objects lie at the same
redshift ($z$\,$=$\,0.236). This and the small projected 
separation of 1$\farcs$8 ($\sim$\,6.7\,kpc) 
suggests that they are in fact interacting. 
This is further supported by recent and ongoing 
star formation in the companion (Sect. \ref{section_companion}).
The images do not show clear morphological
peculiarities or tidal tails but 
both galaxies are embedded in weak
extended emission which is best seen 
in the $R$--band image (Fig.\,\ref{acq_with_contours}). 
In order to assess the amount of extended emission we subtracted two
gaussian point sources, one for the QSO and one for the companion. 
The gaussian width was determined from nearby stars and fits their 
profiles as well as that of the QSO flux peak. 
Table\,\ref{photometry} lists the resulting
photometry for the entire system as well as the flux ratios
QSO/companion and nuclear/total. 
Our analysis shows that the contribution of the extended emission is
largest in $R$ and $J$.

In the following discussion, the name ``J2341'' will refer to
the quasar and the secondary object is called ``companion''.

\subsection{The quasar}
\label{section_quasar}

The spectrum of J2341 shows broad Balmer emission lines
(FWHM $\sim2800$ km/s, rest frame) and a red optical continuum 
(Fig.\,\ref{vlt_ntt}). 
This is unusual for a type--1 AGN. 
We just note that some unusual broad absorption line QSOs (BALQSOs) show very
red SEDs due to wide, overlapping BAL troughs in the ultraviolet (e.g. Hall et
al. \cite{hall02}; Meusinger et al. \cite{meusinger05}). While the continuum
of J2341 is smooth and does not exhibit any indications for BAL structures
over the wavelength range covered by our spectra, the presence of BALs at
$\lambda<3000$\,\AA~cannot be excluded. It is well known that BALQSOs tend to
be more dust reddened than non--BALQSOs (Reichard et al. \cite{reichard03}).

Figure \ref{09447_plus_SDSS} displays the mean SDSS quasar template
 (Vanden Berk et al. \cite{vanden01}) 
over the spectrum of J2341. While J2341 agrees with the optical
quasar template at $\lambda>5500\,\AA$,  it is redder shortwards of
that wavelength where the continuum of classical quasars is dominated
by blue, power--law like emission.  J2341 is also redder
than reddened quasars usually found using optical multi--colour selection
techniques (e.g. Richards et al. \cite{richards03}). The flux ratio of
the broad Balmer emission lines is larger for J2341 than for the
template, indicating that the BLR of J2341 is also reddened. We will
address the reddening further in
 Section\,\ref{section_dust_reddening}. 

\begin{figure}
  \resizebox{\hsize}{!}{\includegraphics[angle=0]{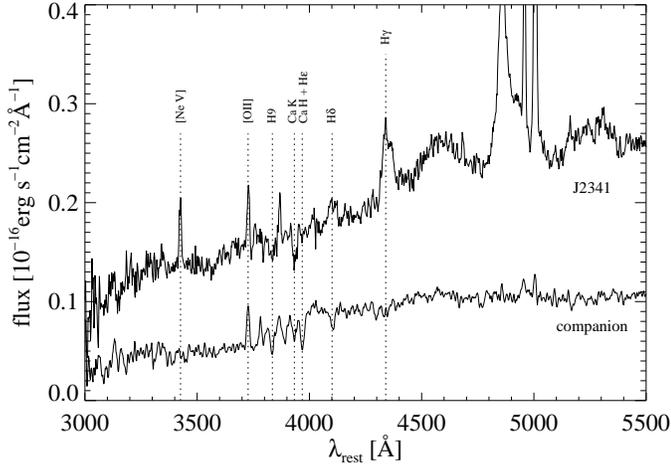}}
  \caption{Blue portion of the spectra of the quasar and its
  companion.
  \label{optical_zoomed}
  }
\end{figure}

\begin{figure}
  \resizebox{\hsize}{!}{\includegraphics[angle=0]{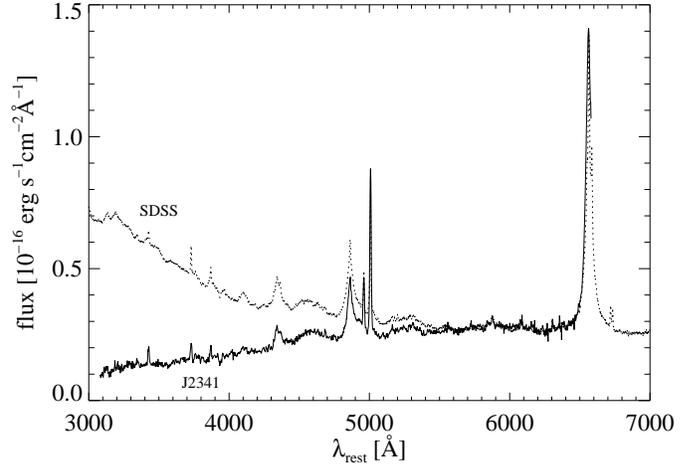}}
  \caption{Spectra of J2341 and the mean SDSS quasar
  template of Vanden Berk et al. (\cite{vanden01}). The SDSS spectrum was
  scaled to match J2341 in H$\alpha$. 
  \label{09447_plus_SDSS}
  }
\end{figure}

Figure \ref{optical_zoomed} shows an enlarged view of the blue portion
of the quasar spectrum. Broad Balmer emission lines and \ion{Fe}{ii} emission 
are detected, as well as low and high ionisation
forbidden lines from the narrow--line region (e.g. [\ion{Ne}{v}],
[\ion{O}{ii}], [\ion{Ne}{iii}], [\ion{O}{iii}]). We also detect Ca II K
as an absorption feature from the host galaxy. 
The presence of Ca II K and the absence
of higher order Balmer absorption suggests that J2341 does not have a
strong contribution from a young--to--intermediate age stellar
population ($<1\,$Gyr). 

In the mid infrared (Fig.\,\ref{mir_spec}) J2341 shows 
a power--law continuum typical for classical AGN (e.g. Weedman
et al. \cite{weedman05}; Ogle et al. \cite{ogle06}; Buchanan et
al. \cite{buchanan06}). We see the [\ion{O}{iv}] emission
line at the 5-$\sigma$ level.
The high ionisation [\ion{Ne}{v}] lines are not detected in the MIR,  
although prominent in the optical spectrum
(Fig.\,\ref{optical_zoomed}).
We also do not detect
the low ionisation emission lines [\ion{Ne}{ii}] and
[\ion{Ne}{iii}]. The detection of [\ion{O}{iv}] and the
upper limits on the [\ion{Ne}{v}] lines are consistent with the fact
that in AGN [\ion{O}{iv}] is usually the brightest MIR line, on
average about three times brighter than the [\ion{Ne}{v}] lines
  (e.g. Haas et al. \cite{haas05}).  
On the other hand, weak PAH emission features at 7.7\,$\mu$m and
11.3\,$\mu$m are seen. Albeit of marginal significance, they are
present at both nod positions of the spectral frames.

The slits of IRS ($\sim$3\farcs6 and $\sim$10\farcs6 for SL and
LL, respectively) include the quasar as well as its companion, 
but the MIR continuum is
likely to be dominated by dust emission from the quasar itself, which is
already 5.5 times brighter at $K_s$ than the companion
(Tab.\,\ref{photometry}).

With $\nu L_{\nu}(15\,\mu m) = 2.1\times10^{44}$\,erg/s the MIR
luminosity of J2341 is relatively low compared to powerful AGN but the
source can still be considered MIR strong (Ogle et
al. \cite{ogle06}). J2341 turns out to be almost a factor of
10 less luminous in the MIR than type\,--2 AGN at similar redshift 
which were found in the ISO--2MASS survey (Leipski et
al. \cite{leipski07}).


It is expected that the interaction of both objects triggers enhanced 
nuclear star formation, when gas and dust are concentrated in the 
central regions.
To study the rate of star formation in quasars, Ho (\cite{ho05})
examined the [\ion{O}{ii}]/[\ion{O}{iii}] flux
ratio of PG quasars. Low ratios between 10\% and 30\% are
consistent with AGN origin. Using this method for J2341 the
low [\ion{O}{ii}]/[\ion{O}{iii}] ratio of $\approx$ 10\% does not
indicate strong ongoing star formation.

In Fig.\,\ref{mir_spec} we overplot the IRS spectrum of the
broad--line AGN/ULIRG Mrk\,231 (Weedman et al. \cite{weedman05}). In
contrast to J2341, Mrk\,231 shows silicate absorption, although broad
permitted lines are present in its optical spectrum. In J2341 on the
other hand it seems that there is not enough dust between the
putative torus (as the source of the MIR continuum emission) 
and the observer to produce significant silicate
absorption. 
Mrk\,231 also 
exhibits a significant FIR excess that indicates powerful hidden
starbursts. J2341 has relatively low FIR emission as derived from
IRAS--ADDSCANs ($F_{60\,\mu{\rm m}}<129$\,mJy) and from the flat
$15-30\,\mu$m slope of the MIR spectrum
(Fig.\,\ref{mir_spec}). Therefore, dusty starbursts as found in Mrk\,231 are 
not present, as already inferred from the optical spectrum. We suggest
that in J2341 the total amount of dust and gas for fuelling starbursts, 
which can be triggered by the interaction, is low.

\subsection{The companion}
\label{section_companion}
The spectrum of the companion is reminiscent of an e(c) galaxy with
rather strong H$\delta$ absorption (Dressler et
al. \cite{dressler99}). As a result of a starburst in the past a
significant population of A type stars exist in the companion as 
traced by prominent Balmer absorption lines (Fig.\,\ref{optical_zoomed}). 
We can speculate that the first
encounter of this galaxy with the quasar has initiated a burst of star
formation in the companion resulting in the strong population of A
stars observed today.

Ongoing star formation is
traced by emission lines like [\ion{O}{ii}] and H$\alpha$. 
This star formation is likely to be triggered by the interaction. 
The observed H$\alpha$ emission could in principle be a contamination 
from an extended narrow--line region of the quasar. However, the low observed 
[\ion{O}{iii}] flux and the very different [\ion{O}{ii}]/[\ion{O}{iii}] 
flux ratio suggests that most of the line emission is intrinsic to the 
companion. 
While the H$\alpha$/H$\beta$ ratio 
appears to be large suggesting significant reddening, the H$\beta$ line 
in this object is affected by strong stellar Balmer absorption 
masking the intrinsic reddening.  

\begin{figure}
  \resizebox{\hsize}{!}{\includegraphics[angle=0]{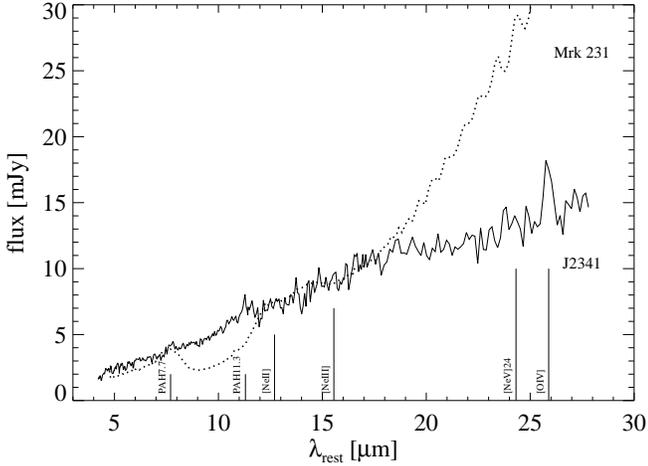}}
  \caption{IRS/Spitzer spectra of J2341 and Mrk\,231 scaled to match J2341 at
    15\,$\mu$m. The positions of important emission lines are marked.  
  \label{mir_spec}
  }
\end{figure}

The absence of strong FIR
emission as evident from the Spitzer spectrum at $20-30\,\mu$m
(Fig.\,\ref{mir_spec}) suggests that there is no strong,
dust--enshrouded starburst in the companion, which could contribute 
to the FIR flux. Since ongoing star formation is weak in
the host galaxy of the quasar but clearly observed in the companion,
most of the PAH features detected in the MIR spectrum
could actually arise in the companion.

\subsection{Comparison with other red quasars}

Our ISPI NIR photometry reveals that for our
slightly extended system some flux is missed when measured via the
(default) point--source photometry of 2MASS
(Tab.\,\ref{photometry}). More appropriate is the 
2MASS aperture photometry (4\arcsec aperture), which includes both
objects. It agrees well with our new photometry
which in turn reduces the $J-K_s$ colour of the quasar alone from 
1.95 to $\sim$\,1.5 (Tab.\,\ref{photometry}). 
The flux ratio of J2341 to its companion is similar in $R$, $J$, and $H$, 
while it significantly different in $K_s$. This indicates that both objects 
are dominated by a comparable old stellar population at red optical to NIR 
wavelengths. In this case their $R-K_s$ colour should also be the same. 
We attribute the 0.6 mag increase in $R-K_s$ in the quasar to 
additional hot dust emission. This also indicates that the central 
regions of the quasar are not severely obscured which otherwise would 
affect the NIR emission. 

The $J-K_s$ colour of J2341 is bluer than the threshold used to select the
2MASS red AGN ($J-K_s>2$; Cutri et al. \cite{cutri02}) which show various
optical continuum slopes (e.g. Smith et al. \cite{smith03})  and X--ray
properties (e.g. Wilkes et al. \cite{wilkes02}). For the 2MASS red AGN we most
likely observe a range from essentially  unobscured  to absorbed
sources. The red $J-K_s$  colour can then be attributed to either clearly 
visible
emission from hot dust in the centre or to absorption of the $J$ continuum.

In the picture where nuclear activity is triggered by interaction red quasars
are supposed to represent young members of the QSO population.  While so far
only a moderate fraction of 2MASS red AGN and classical blue PG
quasars show signs for interaction on HST images ($\sim30\%$; Marble et
al. \cite{marble03}), J2341 clearly has a spectroscopically confirmed
companion (Fig.\,\ref{acq_with_spectra}).


\subsection{On the origin of the red quasar continuum}

We now discuss the faint optical continuum emission of J2341,
  which is unusual for optically selected type-1 QSOs.
The aim is to see how far common mechanisms are able to
  explain the nature of J2341. 
  
First, we assume that J2341 has an intrinsically blue continuum
similar to the SDSS quasar template from Vanden Berk et
al. (\cite{vanden01}), but that J2341 is reddened by a 
dust screen (Sect.\,\ref{section_dust_reddening})
and in addition diluted by stellar light of the host galaxy
(Sect.\,\ref{section_dilution_by_host}).

It should be noted that the SDSS template is quite similar to other
QSO composites like those from the Large Bright Quasar Survey or the
First Bright Quasar Survey (Fig. 11 in Vanden Berk et
al. \cite{vanden01}).
The relative spectrum--to--spectrum variation of the quasars forming the 
SDSS template is about 15--20\% in the wavelength range
3000--5000\,$\AA$ (Fig. 10 in Vanden Berk et
al. \cite{vanden01}), hence relatively small.
The power--law slope
between Ly$\alpha$ and H$\beta$ is somewhat redder
for a composite formed from $z<0.84$ quasars ($\alpha$$_{\nu} = -0.65$; 
$S$\,$\sim$\,$\nu^{\,\alpha}$)
than the slope for the entire quasar sample
entering the SDSS template ($\alpha$$_{\nu} = -0.44$).
But this effect is small for the wavelength range considered here, 
i.e. compared with template from the entire sample, the redder
slope of the $z<0.84$ template reduces the 3000\,$\AA$ continuum by only 
$\sim$10\%, when keeping the 4863\,$\AA$ flux fixed.  
Therefore, we here use the SDSS template from the entire
sample, assuming that it
represents an essentially unreddened quasar spectrum. 
Furthermore, the SDSS template
still contains contributions from the host
galaxy (especially at red optical wavelengths), 
hence for a reasonable comparison any 
host contributions in J2341 will rather tend to make the 
differences to the SDSS template smaller than larger. 

Since these two mechanisms -- dust screen reddening and dilution by
the host galaxy -- 
do not fully explain the red nature of
J2341, we discuss further scenarios in
Section\,\ref{section_dust_geometry}.

\begin{table}
\begin{minipage}[t]{\columnwidth}
\caption{Photometry of the J2341 system, given in Vega magnitudes. The
errors of the NIR photometry are $\le 0.1$ mag. (2) Default 2MASS PSF
photometry includes both objects. The $R$--mag is taken from
USNO--B. (3) 2MASS 4\arcsec aperture photometry. (4) ISPI aperture
photometry, $R$--mag taken from the VLT acquisition frame. (5) Peak
flux ratio of quasar-to-companion from nuclear (1\arcsec) gaussian
fitting. (6) Fraction of total flux contained in nuclear gaussian fits
(quasar + companion); the complementary flux is due to extended emission.}  
\label{photometry}     
\centering                     
\renewcommand{\footnoterule}{}  
\begin{tabular}{cccccc}       
\hline\hline                
Filter & 2MASS & 2MASS & ISPI & QSO/comp & nuclear/total\\  
(1) & (2) & (3) & (4) & (5) & (6) \\
\hline  				        
$R$   & 17.52  & --     & 18.02 & 3.0 & 0.34 \\
$J$   & 16.505 & 16.065 & 16.01 & 3.1 & 0.35 \\
$H$   & 15.502 & 15.234 & 15.21 & 3.3 & 0.42 \\
$K_s$ & 14.555 & 14.577 & 14.49 & 5.5 & 0.44 \\ 
\hline								
\end{tabular}	
\end{minipage}
\end{table}	

\subsubsection{Dust screen reddening}
\label{section_dust_reddening}

In numerous cases it has been shown that the spectra of
luminous red quasars
can be matched with template spectra quite well if simple
dust screen reddening is applied 
(e.g. quasars from the FIRST-2MASS survey, Gregg et al. \cite{gregg02}).
This technique works also well for low--$z$ AGN where host galaxy 
contributions become important. 
For example, 
one nearby red type--1 QSO from the
Spitzer First Look Survey (SSTXFLS\,J171335.1+584756 at $z$\,$=$\,$0.133$; 
Lacy et al. \cite{lacy07}) shares some basic
properties with J2341: FLS\,J1713 has a red optical continuum, 
broad Balmer emission lines (although narrower than in J2341), and Ca II 
absorption lines indicating
contributions from the host galaxy.
However, between 24\,$\mu$m and 70\,$\mu$m the SED of FLS\,J1713 looks more
like that of the ULIRG-AGN Mrk231, with a 70\,$\mu$m\,/\,24\,$\mu$m 
flux ratio of about 4. 
As a test we reddened the SDSS template
using SMC dust extinction curves (Gordon et al. \cite{gordon03}) 
to match the spectrum of FLS\,J1713. 
The rest--frame continuum is decently represented between $\sim$\,$4000\,\AA$
and  
$8000\,\AA$ by the SDSS template reddened with E$_{B-V}=0.5$ (shortwards of 
$\lambda$$_{rest}$ = $4000\,\AA$ no spectral data are available). 
This test demonstrates that even some nearby red quasars (which include 
host galaxy emission) can be explained by
simple screen reddening of optical QSO templates, at least longward of
$\lambda$$_{rest}$ = $4000\,\AA$. 


Figure\,\ref{09447_plus_SDSS} shows the 
spectrum of J2341 and the unreddened SDSS QSO template.
The spectra are scaled to match in the H$\alpha$ line.
This highlights the difference in the
optical continua of J2341 and the template. It 
suggests reddening in J2341 bluewards of $5500\,\AA$ which becomes
particularly strong for $\lambda_{rest}<4000\,\AA$.

\begin{figure}
   \includegraphics[width=8.5cm]{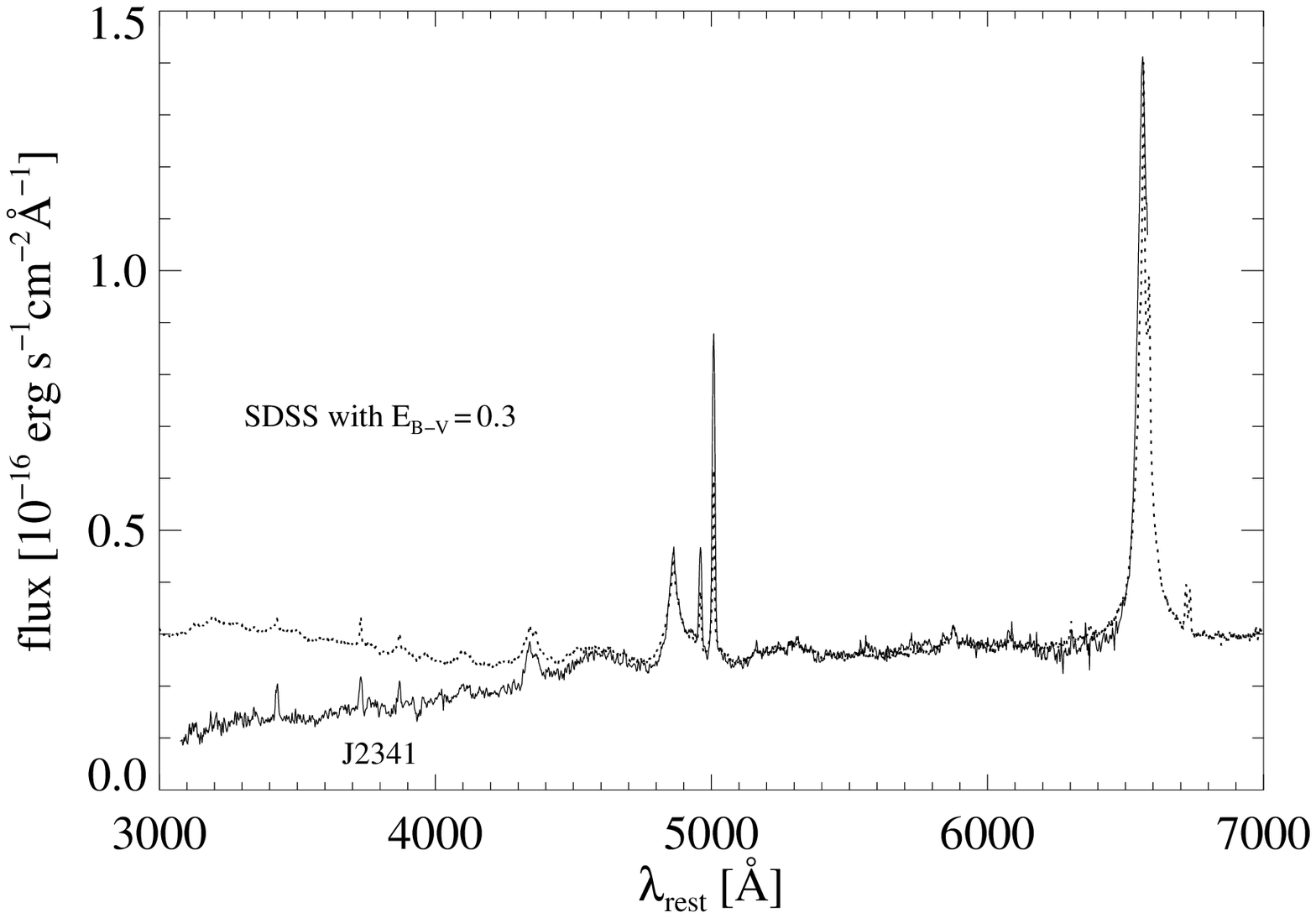}
   \includegraphics[width=8.5cm]{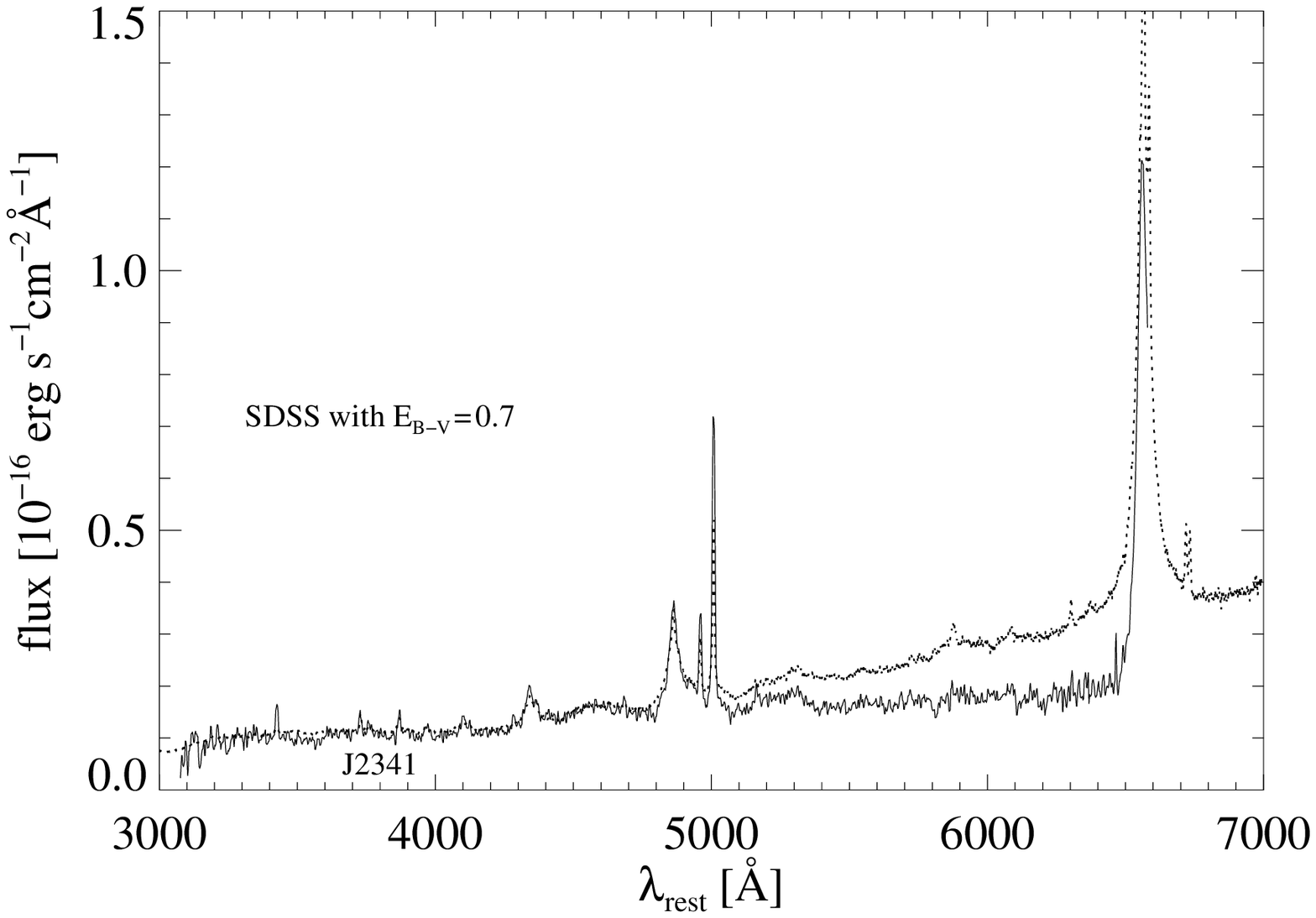} 
      \caption{Comparison of 
	the spectrum of J2341 with
        the SDSS quasar template moderately reddened E$_{B-V}=0.3$
	(top), and strongly reddened E$_{B-V}=0.7$
	(bottom). 
      }
      \label{sequence}
\end{figure}

After applying a dust screen reddening of E$_{B-V}=0.3$ to the template 
both spectra match well down to $\sim$\,$4500\,\AA$ 
(Fig.\,\ref{sequence}, {\it top}).
Also, the fluxes of the broad H$\alpha$ and H$\beta$ lines 
match for both spectra. This shows that a dust screen of 
E$_{B-V}=0.3$ accounts pretty well for reddening on sightlines towards 
the broad--line region of J2341 (assuming the same intrinsic Balmer ratio). 
We determined the uncertainty of the spectral matching to $\Delta E_{B-V}\sim 0.02$. 

However, shortwards of $4500\,\AA$ J2341 still lacks
significant continuum flux compared to the reddened template  
(Fig.\,\ref{sequence}, {\it top}).
At $3000\,\AA$ the discrepancy is about a factor 3. 
An E$_{B-V}=0.7$ would be necessary to
match the blue part of the template with the observed J2341 spectrum 
(Fig.\,\ref{sequence}, {\it bottom}). 
But with E$_{B-V}$\,$=$\,$0.3$ we already have accounted 
for the amount of dust towards the centre as traced by the broad
emission lines.
If a reddening of E$_{B-V}$\,$=$\,$0.7$ was indeed present towards the central regions of 
J2341 the intrinsic H$\alpha$\,/\,H$\beta$ would be $\sim2$ which is not supported 
by either theory (e.g. Kwan \cite{kwan84}) or observations (e.g. Shang et al. \cite{shang07}).
Therefore,
other explanations than extinction by a dust screen have to be considered to account
for the remaining difference at blue optical wavelengths.

\subsubsection{Dilution by host galaxy light}
\label{section_dilution_by_host}

An underlying host galaxy with a pronounced 4000\,\AA~break can
in principle dilute the signature of the blue bump emission, i.e. 
the rising blue continuum. 

We note that QSO templates -- especially
at red optical wavelengths -- are dominated by emission from local, lower
luminosity AGN and, thus, a significant host galaxy component may already
be included  (e.g. Vanden Berk et al. \cite{vanden01}; Gaskell et
al. \cite{gaskell04}). On the other hand, for blue optical wavelengths higher redshift
(and higher luminosity) sources dominate the template and the influence of the
host galaxy on the overall spectral shape of the template is reduced. 

Since the subtraction of several template galaxy spectra taken from the
literature (e.g Calzetti et al. \cite{calzetti94}; Kinney et
al. \cite{kinney96}) give poor results we therefore extracted the host galaxy spectrum directly
from our two dimensional spectral frames (see also Bennert et
al. \cite{bennert06}). This results in a much better fit to
the absorption features in the QSO, even considering the moderate S/N of the
host spectrum. 



\begin{figure}
  \includegraphics[width=8.5cm]{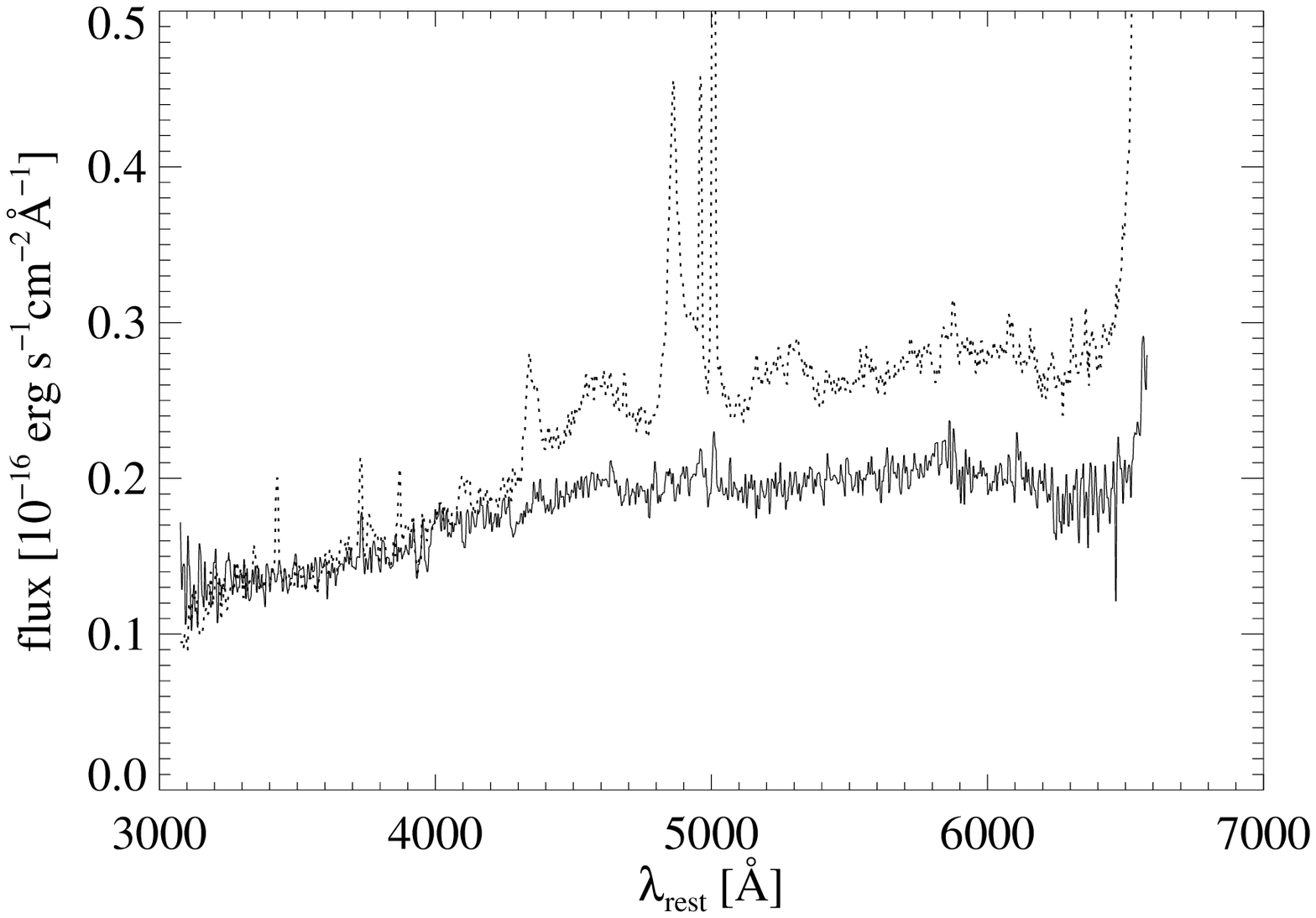}
  \includegraphics[width=8.5cm]{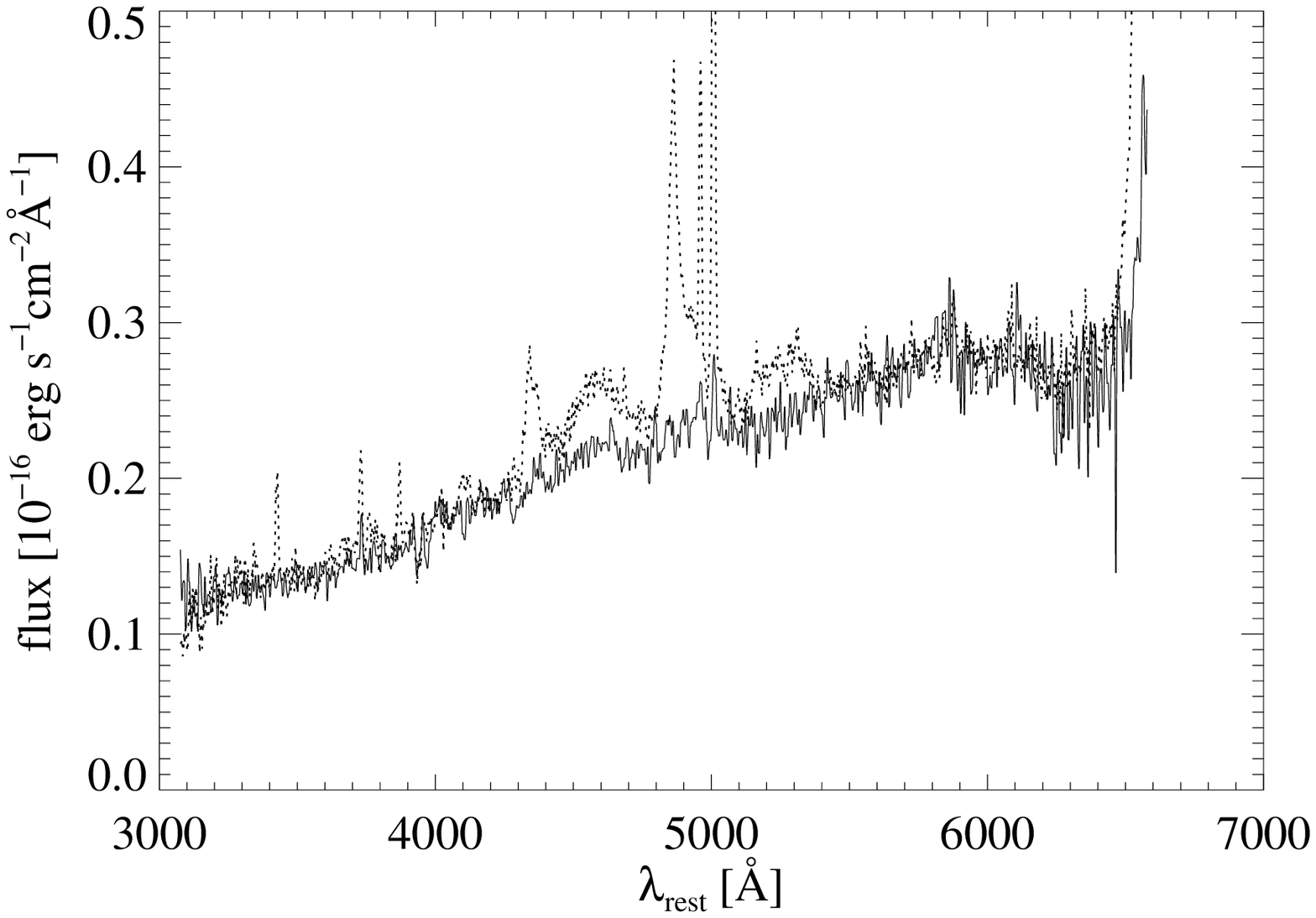}
   \caption{The extracted host spectrum without (top) and with additional
               reddening of E$_{B-V}$\,$=$\,$0.3$ ontop of the QSO spectrum (dashed line). The host
               spectra are scaled to remove the Ca II K
               feature from the QSO after subtraction. Just for these
               graphs additional offsets were applied to the host spectra to
               facilitate an easy comparison with the QSO.
   }
         \label{host}
\end{figure}

The host  spectrum was extracted from an area of 0\farcs8 with a separation of
1\arcsec~south--east (3.7\,kpc) to the region of the QSO spectrum
(Fig.\,\ref{acq_with_contours}).  The placement and width of the region used
for the host spectrum extraction was chosen to provide maximal S/N without
contamination from nuclear QSO light.  The host spectrum shows clear Ca II
H+K absorption as well as [\ion{O}{ii}] and narrow H$\alpha$/[\ion{N}{ii}]
emission (Fig.\,\ref{host}, {\it top}).  For the same reasons as for the
companion spectrum we expect the stellar template to be largely free of 
AGN contamination (no broad--line components, no high--ionisation lines; see \S\,3). 
While the [\ion{O}{ii}], H$\alpha$, and the weak
[\ion{O}{iii}] emission could result from an extended narrow--line region, it
can also indicate ongoing star formation in the host galaxy. The flux ratio of
F$_{\rm [\ion{O}{ii}]}\geq{\rm F}_{\rm [\ion{O}{iii}]}$ suggests star
formation as cause for the line emission in the host galaxy.


Following Boroson et al. (\cite{boroson82}) we also obtained a host galaxy 
spectrum by subtracting the nuclear 0\farcs6 spectrum from a spectrum of 
3\farcs0 width which covered the quasar plus host emission (on the opposite 
side from the companion). The nuclear 
spectrum was scaled in such a way that the broad lines were removed and only 
the underlying host galaxy spectrum remained. This spectrum agrees very 
well with the spectrum from the off--nuclear extraction supporting our 
approach. In the further analysis only the off--nuclear spectrum was used.
 
We scaled the host spectrum in such a way that, after subtraction from the QSO
spectrum, the Ca II K line is removed. We also tried other scalings but 
even the most extreme
values explored were not able to recover the blue bump emission while
on the other hand highly overestimating the Ca II K absorption. 
Thus, we used the Ca II K line as a measure for the host galaxy contribution.

\begin{figure}
   \includegraphics[width=8.5cm]{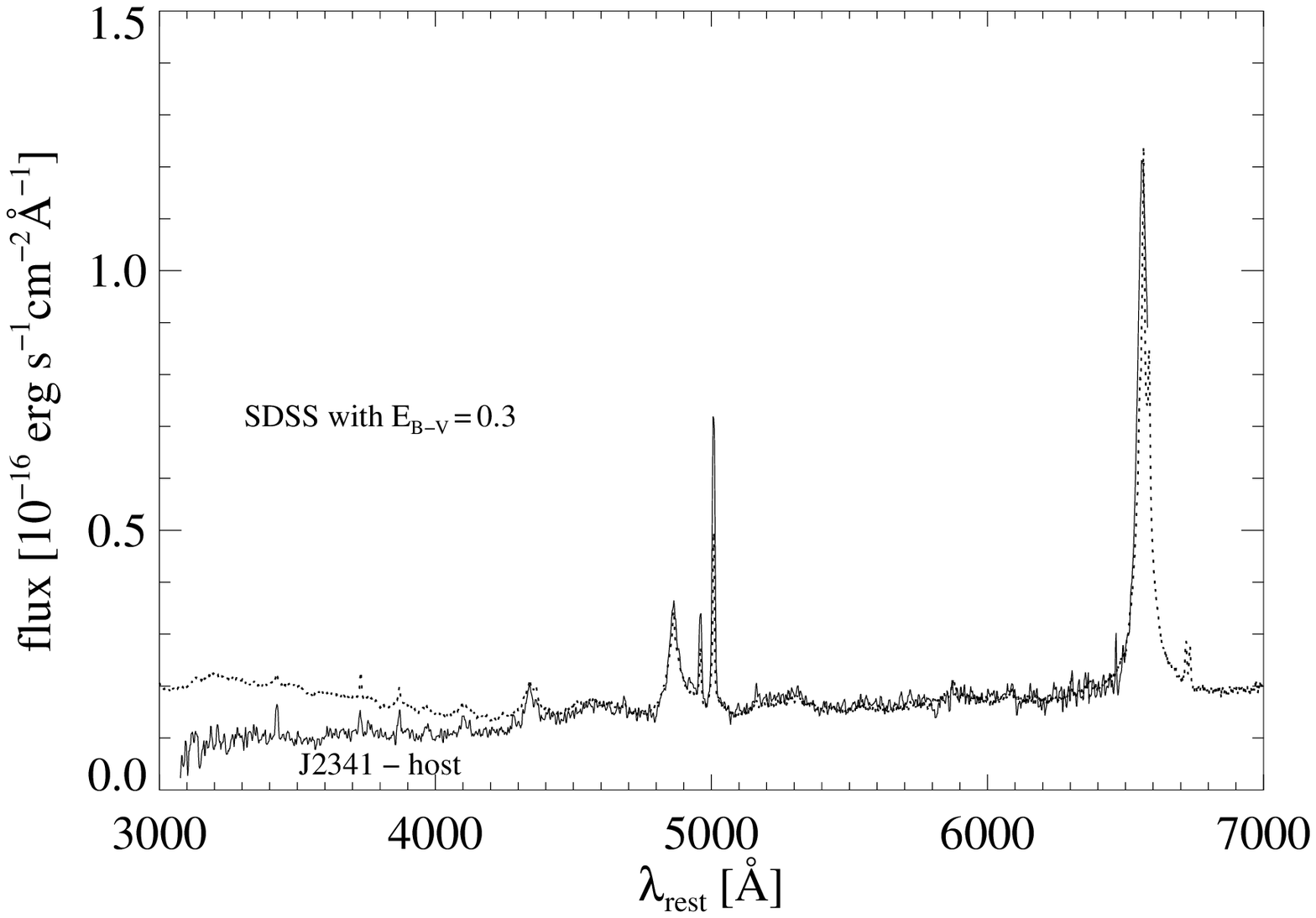}
   \includegraphics[width=8.5cm]{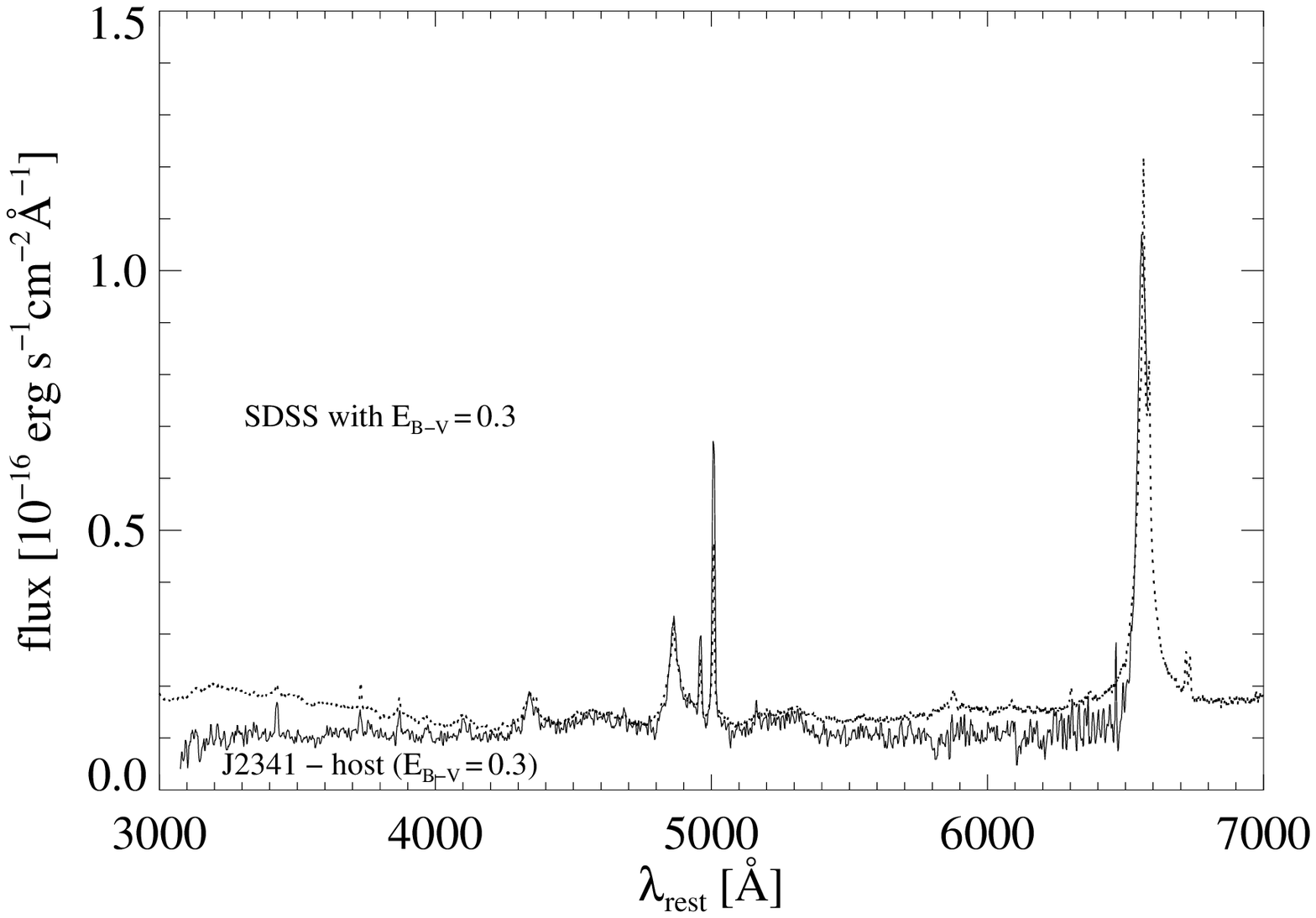} 
      \caption{Comparison of the moderately reddened SDSS quasar template with
	the spectrum of J2341 after subtraction of the host without reddening
	(top), and with additional reddening 
	(bottom). 
      }
      \label{dilution_by_host}
\end{figure}

This subtraction also recovers the
H$\epsilon$ line in a reasonable strength (Fig.\,\ref{dilution_by_host}, {\it
top}). This line was largely absent in the original spectrum due to 
the underlying Ca II H absorption. Down to about 4500\,$\AA$
the resulting ``host--free'' QSO spectrum of J2341
fits the dust reddened SDSS QSO template better than the original
J2341 spectrum does. But shortwards of 4500\,$\AA$ the ``host--free'' 
J2341 system still lacks continuum, and at 3000\,$\AA$ the discrepancy is
about a factor of 2.
This is too large to be explained by
the 10--20\% spectrum-to-spectrum variation of the SDSS template or
the 10\% redder slope of the $z<0.84$ template. 
Even if a redder template made of extraordinarily UV--weak quasars would fit
J2341 better, it is of particular interest to explore the nature of
such potentially UV--weak quasars.

A refined way to account for the remaining discrepancies between the QSO
template and the J2341 spectrum would be to consider increased 
reddening towards the centre of the system, as found by
e.g. Bennert et al. (\cite{bennert06}) for their Seyfert galaxies.
Then the nuclear host galaxy component, as seen in the J2341 QSO
spectrum, would be more reddened than the extracted host galaxy
template 3.7\,kpc further out.
Reddening of the host galaxy spectrum before
subtraction from the QSO gives an interesting results 
(Fig.\,\ref{host}, {\it bottom}): After applying a redding
equivalent to E$_{B-V}=0.3$ the reddened host galaxy spectrum follows the
spectral shape of the QSO remarkably closely. (We chose E$_{B-V}=0.3$ since
that was the value inferred as reddening towards the broad--line region.) Due
to the similarity in spectral shape the
subtraction of the reddened host galaxy spectrum from the QSO results in the
QSO spectrum being almost flat in flux over the observed wavelengths 
(Fig.\,\ref{dilution_by_host}, {\it bottom}). This makes the fit of the J2341
spectrum with the SDSS QSO template worse over almost the whole wavelength
range compared to the case with an unreddened host galaxy.

It is also possible that the stellar population in the centre
differs from that at a few kiloparsec further out. But
Bennert et al. (\cite{bennert06}) found this issue to be only of minor
importance for their spatially well resolved object (NGC\,1386).

To summarise, so
far the comparison with the SDSS quasar template 
strongly suggest that dust screen reddening alone and the combination of
screen reddening with host galaxy
contributions are not able to fully explain the red optical/UV
continuum of J2341 shortwards of H$\beta$. 
Therefore, we consider alternative scenarios.

\subsubsection{Refined scenarios and alternatives}
\label{section_dust_geometry}

In this section we discuss special dust distributions and other
effects. 
So far we have applied a dust screen of E$_{B-V}=0.3$ to the SDSS
template in order to match its spectrum with J2341 after subtraction
of the host.
The dust cannot be distributed evenly throughout the whole galaxy,
because we clearly detect features of the host galaxy in the optical spectrum
of the quasar, while the AGN continuum appears heavily suppressed
(Fig.\,\ref{optical_zoomed}).  Moreover, the emission lines from the
NLR  are still clearly visible in J2341 suggesting that the major part
of the obscuring material is located in the centre of
the galaxy, somewhere between the NLR and the accretion disc 
(e.g. Fig.\,\ref{optical_zoomed}).

Small, sharply confined dust clouds with a low volume filling factor
may lie by chance in our line of sight towards the accretion disk, 
which produces the continuum emission.
Then the clouds  may redden the AGN continuum emission,
while the bulk of the slightly more extended BLR may be less affected.
This scenario appears in accordance with results from X--ray
spectroscopy of red 2MASS type--\,1 AGN, where the presence of a
compact absorber very near to the active nucleus has been proposed.
In these AGN the BLR is clearly visible, but the X--ray
source is significantly obscured (e.g. Wilkes et al. \cite{wilkes05}). We
should note, however, that the X--ray absorption traces the gas while the 
optical obscuration is due to dust.
In addition, if a low volume filling factor is assumed, 
the swirling dust clouds are expected to shield the centre
not permanently (transient shielding). Hence, we expect significant
variability in both the total intensity and the spectral
shape. But we detect
neither a significant difference in the spectral slope of our spectra
(that span two years) nor are significant photometric differences
observed by comparison of SDSS and USNO--B data (that span 40 years).

Another explanation considers that light from the nucleus grazes the
edge of a moderately inclined torus so that the BLR is largely seen, but
the accretion disk is essentially hidden and strongly reddened (see 
also Smith et al. (\cite{smith04}) who proposed a similar scenario to 
explain the polarisation properties of a group of Seyfert 1 galaxies). Then the torus must have a
rather sharp edge, a requirement which looks unlikely because the
ongoing interaction favours a more disturbed
dust distribution. In addition, the MIR spectra do not show recognisable 
silicate absorption placing an upper limit on the MIR absorption and, 
thus, on the optical absorption of A$_{\rm V}$ $<$ 1 (following Kr\"ugel et
al. \cite{kruegel03}). This upper limit is consistent with our finding
of E$_{B-V}=0.3$. Therefore even this special geometry does not provide 
a satisfying explanation for substantially higher extinction towards
the accretion disk. 

Next we discuss screen versus mixed case reddening. 
Simple screen reddening by dust is not suitable to explain simultaneously the
reddening of the BLR and the continuum. But dust mixed with the
BLR may be in accordance with the observations. Then the BLR
suffers only moderate mixed case reddening, while the continuum is
more strongly reddened by the entire 
absorbing screen. However, this requires dust inside the
dust sublimation radius mixed with the BLR, where it should ultimately be 
destroyed by the strong radiation field of the nucleus.

Since common mechanisms fail to provide a satisfying explanation for
 the red continuum of J2341, it seems
 that the strong Blue Bump emission of average quasars is
 largely missing in J2341.
 For low--luminosity AGN (LLAGN) radiatively inefficient accretion
 flows (RIAFs) were proposed to explain the lack of prominent Big Blue Bump
 emission in the UV (e.g. Ho \cite{ho99}; Narayan
 \cite{narayan05}). Although J2341 qualifies as a bona fide quasar from
 its luminosities at different wavelengths,
 the spectral shape towards the UV is reminiscent
 of RIAFs. Since the dust models have difficulties to consistently
 explain the spectrum of J2341, the  RIAF scenario is an attractive
 possibility to interpret the observations. However, the detection of
 prominent high excitation lines like [\ion{Ne}{v}] strongly questions an
 intrinsically low UV flux.

Even the refined scenarios have difficulties matching
 J2341 with standard blue quasar templates. 
 This leads us to suggest that the continuum of the quasar J2341
 is intrinsically redder than that of typical quasars, 
 although not being in a RIAF phase.

  Since the enigmatic UV--weak nature of J2341 may also be found in other
  objects further studies of this prototypical object are desired.

 Higher S/N data in which several stellar absorption features are securely
 identified could allow for a more detailed way to account for the host 
 galaxy contribution. On the other hand, observations in the rest--frame 
 UV range would be especially interesting for the apparent lack of a 
 prominent Small Blue Bump while strong Balmer and \ion{Fe}{ii} emission 
 is observed in the optical. Studies sampling this wavelength range are 
 needed to further address this phenomenon.

\begin{acknowledgements}
  Part of this work was supported by Sonderforschungsbereich SFB\,591
  ``Universelles Verhalten gleich-gewichtsferner Plasmen'' der
  Deutschen Forschungsgemeinschaft, and by Nordrhein--Westf\"alische
  Akademie der Wissenschaften.  We thank Dan Weedman for kindly
  providing the Spitzer spectrum of Mrk 231.  We also thank Andy
  Robinson for his helpful comments on the manuscript.
  Based on observations with the Infrared Space Observatory ISO, an
  ESA project with instruments funded by ESA Member States (especially
  the PI countries: France, Germany, the Netherlands and UK) and with
  the participation of ISAS and NASA. The Two Micron All Sky Survey is
  a joint project of the University of Massachusetts and IPAC/Caltech,
  funded by the National Aeronautics and Space Administration and the
  National Science Foundation.

\end{acknowledgements}

\end{document}